\documentclass[11pt]{article}
 \usepackage{amsfonts}
 \usepackage{amsmath} 
 \usepackage{amssymb}
 \newfont{\bbbold}{msbm10 scaled \magstep1}

 \def\bbA{\mbox{\bbbold A}}

 \def\cD{{\cal D}}
 
 \def\cF{{\cal F}}

 \def\cM{{\cal M}}

 \newfont{\goth}{eufm10 scaled \magstep1}

 \def\a{\alpha}
 \def\b{\beta}
 \def\c{\gamma}
 \def\d{\delta}
 \def\e{\epsilon}

 \def\m{\mu}

 \def\t{\tau}
 \def\th{\theta}

 \def\del{\partial}
 \def\ua{\underline{\alpha}}

 \def\una{\underline a}\def\unA{\underline A}
 \def\unb{\underline b}
 \def\unC{\underline C}

 \def\unM{\underline M}

 \def\unH{\underline{H}}\def\unG{\underline{G}}

 \def\xz{\times}

 \def\nab{\nabla}
 
 \def\tpsi{\tilde{\psi}} 
 \def\tL{\tilde{L}}
 \def\tJ{\tilde{J}}

 \def\del{\partial}

 \def\be{\begin{equation}}\def\ee{\end{equation}}
 \def\bea{\begin{eqnarray}}\def\eea{\end{eqnarray}}
 \def\ba{\begin{array}}\def\ea{\end{array}}

 \newcommand{\eq}[1]{(\ref{#1})}
 
 \def\det{{\rm det\,}}
 \def\Tr{{\rm Tr}}
 \def\Str{{\rm Str}}
 \def\One{1 \!\! 1}
 \def\hF{\hat{F}}
 
\begin{document}
 \newcommand{\hoch}[1]{$\, ^{#1}$}
\newcommand{\uppsala}{\it\small Department of Theoretical Physics, 
Uppsala University, Uppsala, Sweden}
\newcommand{\newton}{\it\small Isaac Newton Institute for Mathematical
Sciences, Cambridge, UK}
\newcommand{\kings}{\it\small Department of Mathematics, King's College,
London, UK}


\newcommand{\auth}{\large J.M. Drummond\hoch{1}, P.S. Howe\hoch{1,3},  and U.
Lindstr\"om \hoch{2,3} }

 \thispagestyle{empty}

 \hfill{KCL-TH-00-14}

 \hfill{UUITP-08/02}

 \hfill{NI02019-MTH}

 \hfill{hep-th/0206148}

 \hfill{\today}

 \vspace{20pt}

 \begin{center}
 {\Large{\bf Kappa-symmetric non-abelian Born-Infeld actions }}

 {\Large{\bf in three dimensions}}
 \vspace{30pt}

\auth

\vspace{15pt}

\begin{itemize}
\item [$^1$] \kings
\item [$^2$] \uppsala
\item [$^3$] \newton
\end{itemize}

\vspace{60pt}

 {\bf Abstract}

 \end{center}

A superembedding construction of general non-abelian
Born-Infeld actions in three dimensions is described. These actions have rigid
target space and local worldvolume supersymmetry (i.e. kappa-symmetry). The standard abelian Born-Infeld gauge multiplet is augmented with an additional
worldvolume $SU(N)$ gauge supermultiplet. It is shown how to construct
single-trace actions and in particular a kappa-supersymmetric extension of the symmetrised trace action.

 {\vfill\leftline{}\vfill \vskip  10pt

 \baselineskip=15pt \pagebreak \setcounter{page}{1}


\section{Introduction}

In this paper we use the superembedding method to give a construction
of non-abelian brane actions of Born-Infeld type. The superembedding
construction guarantees that these actions have manifest local
worldvolume supersymmetry (or kappa-symmetry) as well as the rigid
supersymmetry of the flat target space. One such action has been
constructed before for the case of coincident D-particles
\cite{sorokin}. In this case the worldvolume gauge field strength
vanishes since the worldvolume is only a one-dimensional worldline.

There has been much discussion in the literature of possible
generalisations of abelian Born-Infeld theory to the non-abelian
case. This has largely been motivated by the fact that the effective action
of coincident D-branes in type II superstring theories is some
non-abelian generalisation of the Born-Infeld action. One proposal for
this generalisation is to keep the form of the abelian action but
replace the $U(1)$ field strengths with $U(N)$ field strengths and
then take a symmetrised trace over the orderings of the different
field strengths  at each order \cite{tseytlin}. This correctly
reproduces the $F^4$ terms calculated from string scattering
amplitudes but  fails at higher orders in the field strengths
\cite{ht,bdrs,stt}. In \cite{ketov} another approach to constructing
non-abelian Born-Infeld actions was considered using a
generalisation of the partially broken global supersymmetry approach
to constructing abelian Born-Infeld actions with linear and non-linear
supersymmetry \cite{Bagger:1996wp,Rocek:1997hi,Ivanov:1999fw}.

A different approach to constructing the non-abelian Born-Infeld action has been developed in refs \cite{Koerber:2001uu,Koerber:2001hk,Koerber:2001ka} while attempts to find it using ten-dimensional supersymmetry have been given in refs \cite{Cederwall:2001td,Cederwall:2001dx,Cederwall:2002df} and in
ref \cite{Collinucci:2002ac}.

Space-filling branes have been considered from the perspective of
superembeddings in \cite{bppst,abkz,codzero,sven}. The Green-Schwarz
action for the space-filling membrane was constructed in \cite{codzero}. 
In section 2 we give a brief summary of this method applied to
the construction of the abelian Born-Infeld action. In section 3
we show how to generalise this action in the presence of a non-abelian
worldvolume $SU(N)$ gauge supermultiplet. By
assumption this multiplet is described by a 2-form, $F$, satisfying the
standard Bianchi identity $DF=0$. We then
consider actions which involve the $U(1)$
and $SU(N)$ fields in such a way that they combine into a single
$U(N)$ field strength. We concentrate on Lagrangians which
have the form a single trace of some function of these field strengths. Our
method is similar to that used in \cite{hl} to construct 
kappa-symmetric higher derivative terms in brane actions. The essential idea
is to construct a closed worldvolume Lagrangian form using the abelian
and non-abelian worldvolume fields.


\section{Abelian action}

In this section we give a brief review of the superembedding
construction of the abelian Born-Infeld action for a supermembrane
in flat $N=2$ three dimensional superspace \cite{codzero}.

\subsection{Superembedding formalism}

We consider a superembedding, $f: \cM \longrightarrow
\underline{\cM}$. For the purposes of this paper the target space,
$\underline{\cM}$, will always be flat. Our index conventions are as
follows; coordinate indices are taken from the middle of the alphabet
with capitals for all, Latin for bosonic and Greek for fermionic,
$M=(m,\m)$, tangent space indices are taken in a similar fashion from
the beginning of the alphabet so that $A=(a,\a)$. The distinguished
tangent space bases are related to coordinate bases by means of the
supervielbein, $E_M{}^A$, and its inverse $E_A{}^M$. Coordinates
are denoted $z^M=(x^m,\th^{\m})$. We use exactly the same notation
for the target space but with all of the indices underlined.
Target space forms are written with an underline, e.g. $\unH$. Their 
pullbacks are written without an underline, $f^{*} \unH = H$.

The embedding matrix is the derivative of $f$ referred to the
preferred tangent frames, thus
\be
E_{A}{}^{\unA} = E_{A}{}^{M} \del_{M} z^{\unM} E_{\unM}{}^{\unA}. \label{embedding matrix}
\ee
This tells us how to pull back target space forms onto the worldvolume,
\be
f^* E^{\unA} = E^A E_{A}{}^{\unA}.
\ee

The basic embedding condition is
\be
E_{\a}{}^{\una} = 0. \label{embedcond}
\ee
Geometrically, this says that at each point on the brane the odd
tangent space of the brane is a subspace of the odd tangent space of
the target space.
In general, this condition gives constraints on the superfields
describing the worldvolume theory. For codimension zero, however, it can
be enforced without loss of generality as discussed in \cite{codzero}.

The worldvolume multiplet is described by the transverse target space
coordinates considered as superfields on the worldvolume. For the
space-filling membrane we embed an $N=1$ superspace into an $N=2$
superspace of the same bosonic dimension. Thus, in the absence of
further constraints, our worldvolume multiplet is an unconstrained
spinor superfield. This superfield, being associated with the breaking
of one of the supersymmetries, is referred to as the Goldstone
superfield.
 
The dimensions of the coordinates $x$ and $\th$ are $-1$ and
$-\tfrac{1}{2}$ respectively while the worldvolume superspace derivatives $\cD_a$, $\cD_{\a}$ have
dimensions $+1$ and $+\tfrac{1}{2}$ respectively. In the superembedding context it is natural to take the bosonic
component of the gauge connection $A_a$ to have dimension $-1$
so that the non-abelian  field strength two-form should be defined by 
$F=dA +\frac{1}{\a'}A^2$ (where $\a'$ has dimension minus two). The purely even component, $F_{ab}$, has dimension zero. 
\subsection{Space-filling membrane}

We now specialise to the membrane in flat three-dimensional
$N=2$ superspace \cite{codzero}. The bosonic indices for the
worldvolume and target space may be identified since we are considering a space-filling
brane. The fermionic target space indices are written $\ua = \a i$
where $i = 1,2$ since we embed an $N=1$ superspace into an $N=2$
superspace. The index $\a$ is a real, two-component Majorana spinor
index. The internal index $i$ is an $SO(2)$ index.

\subsubsection*{Background forms}

The Neveu-Schwarz 3-form field strength $\unH$ and the Ramond field
strengths $\unG_2$ and $\unG_4$ satisfy the Bianchi identities,
\be
d\unH = 0 ,
\hspace{20pt}
d\unG_2 = 0 ,
\hspace{20pt}
d\unG_4 = \unG_2 \unH.
\ee
A solution to these equations in flat space is given by forms
whose non-zero components are
\be
H_{\a i \b j c} = -i(\c_c)_{\a \b}(\t_1)_{ij},
\hspace{20pt}
G_{\a i \b j} = -i\e_{\a \b}\e_{ij},
\hspace{20pt}
G_{\a i \b j cd} = -i(\c_{cd})_{\a \b} (\t_3)_{ij}.
\ee

The field strengths $\unG_2$ and $\unG_4$ are related to the potentials $\unC_1$ and $\unC_3$ by
\be
\unG_2  = d \unC_1,
\hspace{20pt}
\unG_4 = d \unC_3 - \unC_1 \unH.
\ee

The components of the Ramond potentials $\unC_1$ and $\unC_3$ depend
on the target space coordinates and can be expressed, in a particular gauge, in terms of
$\th^{\a 2}$ only. When the brane is embedded into the $N=2$ target superspace $\th^{\a 2}$ becomes the Goldstone superfield in the static gauge.

\subsubsection*{Worldvolume Supergeometry}

We parametrise the odd-odd part of the
embedding matrix as follows
\be
E_{\a}{}^{\b 1} = \d_{\a}^{\b}
\hspace{10pt}
\text{and}
\hspace{10pt}
E_{\a}{}^{\b 2} = h_{\a}{}^{\b}.
\ee

where $h_{\a \b} = k\e_{\a \b} + h_a(\c^a)_{\a \b}$
In addition we can set the even-even part of the embedding
matrix to be trivial, $E_{a}{}^{\unb} = \d_{a}{}^{b}$, since the bosonic
dimensions of the brane and the target space are the same. We denote the worldvolume superspace derivatives in the embedding basis by

\be
\cD_A=E_A{}^M\del_M
\ee

We can now calculate the worldvolume torsion by pulling back the standard
flat target space torsion. As noted previously for the case of space-filling
branes, we do not have to introduce a worldvolume connection
\cite{codzero,bppst,abkz} so that

\be
[\cD_A,\cD_B]=-T_{AB}{}^C\cD_C\,.
\ee

The dimension zero component of the
worldvolume torsion is

\be
T_{\a \b}{}^a = -i(\c^b)_{\a\b}m_{b}{}^{a}, \label{WVtorsion}
\ee
where
\be
m_{ab} = (1 + k^2 + h^2)\eta_{ab} - 2h_a h_b -2 \e_{abc} k h^c.
\label{mmatrix} 
\ee

The other components of $T_{AB}{}^C$ can be found straightforwardly but we shall not need them in this paper.

To describe the worldvolume multiplet we introduce a worldvolume
2-form $\cF$ (the modified field strength). This is 
satisfies the Bianchi identity 
\be
d\cF = -H \label{wvcfbianchi}, 
\ee
where $H$ is the pullback of $\unH$ onto the worldvolume. To get the 
required worldvolume $N=1$ Maxwell multiplet we impose the standard
$\cF$-constraint $\cF_{\a \b} = \cF_{\a b} = 0$. The constraint
$\cF_{\a \b} = 0$ tells us that we have an $N=1$ Maxwell multiplet on
the brane as well as the Goldstone fermion of the embedding, while the
constraint $\cF_{\a b} = 0$ eliminates one of these spinor superfields
in terms of the other. This leaves us with just the degrees of freedom
associated with the (off-shell) Maxwell multiplet. The Bianchi identity then gives a
formula for $\cF_{ab}$ in terms of the degrees of freedom of the
embedding. We find that $k = 0$  and $\cF_{ab} = \e_{abc}\cF^c$, where
\be
\cF_a = \frac{2h_a}{1 + h^2} \label{relhtoF}
\ee
The $h_{\a}{}^{\b}$ field in the embedding matrix is therefore related
to the field strength tensor of the Maxwell multiplet in a non-linear
fashion.

\subsubsection*{Action}

To construct an action we start with the closed Wess-Zumino 4-form
\cite{hrs},
defined on the worldvolume by $W_4 = G_4 + G_2 \cF$, where $G_2 =
f^*\unG_2$ and $G_4 = f^*\unG_4$. By construction we can
write the Wess-Zumino form explicitly as $W_4 = dZ_3$ where $Z_3 = C_3
+ C_1\cF$ and $C_1 = f^*\unC_1$, $C_3 = f^*\unC_3$. The components of
$Z_3$ depend explicitly on the target space coordinates.

Since $W_4$ is a form of degree one higher than the bosonic dimension
of the worldvolume, the fact that it is closed implies it is exact and
we can write $W_4 = dK_3$. The components of $K_3$ do not explicitly
depend on the target space coordinates. The Lagrangian form for the
abelian action is then
\be
L_{3}^{U(1)}=K_3-Z_3,
\ee
and is closed by construction. We can solve $W_4 = dK_3$
for $K_3$. We find its only non-zero component is given by
\be
K_{abc} = \e_{abc}K,
\hspace{10pt}
\text{where}
\hspace{10pt}
K = \frac{1 - h^2}{1 + h^2}.
\ee
Given the relation (\ref{relhtoF}) between $h_a$ and $\cF_a$ we find
that $K$ has the standard Born-Infeld form,
\be
K  = \sqrt{ - \det (\eta_{ab} +
\cF_{ab})}= \sqrt{1 - \cF^2}, \label{BIterm}
\ee

where $\cF^2:=\cF^a\cF_a$.
The top component of the Lagrangian form is 
$L_{abc}^{U(1)} = \e_{abc}L^{U(1)}$ where
$L^{U(1)}=(K-Z)$. The first term, $K$, is the
Born-Infeld part and the second, $Z$, is the Wess-Zumino
term. We convert into the coordinate basis using the even-even component of the worldvolume
vielbein $E_{m}{}^{a}$.

Finally, the Green-Schwarz action for the brane is defined by
\be
S^{U(1)} = \int d^3x \hspace{3pt} (\det E)L^{U(1)}. \label{abelianS}
\ee
The superfields in the integrand are evaluated at $\th = 0$.
The closure of $L_{3}^{U(1)}$ ensures that this action is invariant under
general diffeomorphisms of the worldvolume. The odd diffeomorphisms
are identified with kappa-symmetry as described in \cite{hrs}. 


\section{Non-abelian actions}
In this section we consider ways of generalising the approach of
the previous section in the presence of non-abelian worldvolume fields.
We start by considering the general structure of such Lagrangians and
then show how to construct non-abelian Born-Infeld
Lagrangians. 

\subsection{General structure of non-abelian actions}

We now discuss possible generalisations to the abelian Born-Infeld
action. We look for Lagrangians which are functions of a non-abelian
field strength, $\hat{F}$, taking values in the Lie algebra of $U(N)$.
We take a similar point of view to \cite{sorokin} whereby we look for
actions which are invariant under a single kappa-symmetry and regard
any other kappa-symmetries to have been gauge-fixed. This is in
contrast to \cite{bergs} where the parameter of the kappa-symmetry is taken to trasform under the adjoint representation of
$U(N)$ adjoint. This adjoint kappa-symmetry was shown to be
inconsistent after a certain order in the field strengths
\cite{bergs2}. Accordingly, we keep the picture of a single brane
embedded in a flat target space but introduce an extra worldvolume
$SU(N)$ field strength, $F$ and construct Lagrangians which are functions of
$\hat{F}=\cF\One+F$.

Our main interest in these actions arises from the
fact that the effective action of coincident D-branes in type II
string theories is some generalisation of the abelian Born-Infeld
action which is invariant under two target space supersymmetries and
one local worldvolume supersymmetry (kappa-symmetry). As such we shall
restrict our attention to actions which take the form of a single
trace of some function of the worldvolume fields.



We recall that in three dimensions we can write the pure $\cF$ terms
of the (super) Born-Infeld Lagrangian as (\ref{BIterm}),
\be
  \sqrt{-\det (\eta_{ab} + \cF_{ab})} = \sqrt{1 - \cF_a \cF^a} 
= \sum_{n=0}^{\infty} c_n \cF_{a_1} \cF^{a_1} .... \cF_{a_n} \cF^{a_n}. 
\ee
This Lagrangian can be defined as the effective Lagrangian given by
open strings ending on a D-brane excluding derivative corrections.
One ambiguity in passing to the non-abelian case is that one can
exchange terms with antisymmetric pairs of covariant derivatives with
terms involving commutators of the field strength, using the relation
$\a^{\prime}\nab_{[a} \nab_{b]} \hF_{cd} \sim [\hF_{ab},\hF_{cd}]$. (The explicit $\a'$ is a consequence of our definition of $F$ in section 2.1.)
We
use the term derivative corrections to refer to terms with derivatives
which cannot be cast into commutator form.
The pure $\hat{F}$ terms in any non-abelian generalisation
can then be written as
\be
\Str \sqrt{\One - \hat{F}_a \hat{F}^a} + \text{ commutator terms} +
\text{ derivative corrections},
\label{NBIpureF}
\ee
where the $U(N)$ field strengths are denoted by
$\hat{F}_{ab}=\e_{abc}\hat{F}^{c}$ and $Str$ denotes the symmetrised trace. 
The symmetrised trace part can be written as
\be
\sum_{n=0}^{\infty} c_n \hat{F}^{a_{1} R_{1}} \hat{F}_{a_{1}}^{R_{2}}....
\hat{F}^{a_{n} R_{2n - 1}} \hat{F}_{a_{n}}^{R_{2n}} D_{R_{1}..R_{2n}},
\ee
where the coefficients $c_n$ are the same as those in the abelian case
and $D_{R_{1}....R_{2n}} = \Str (T_{R_{1}}....T_{R_{1}})$.

Any terms with commutators, including any pure $\hat{F}$ terms of odd
powers, vanish in the abelian limit, while the symmetrised trace part
reduces to the usual abelian Lagrangian.

\subsection{Superembedding construction of non-abelian actions}

To construct actions of the type discussed above, we use a similar
method to that used to construct kappa symmetric higher derivative
terms in brane actions \cite{hl}. We introduce non-abelian fields
onto the worldvolume and construct another closed Lagrangian 3-form
$L_3$ out of these fields by specifying its lower-dimensional
components and solving $dL_3 = 0$ \cite{Gates:1997kr,
Gates:1997ag}. 

The $U(1)$ multiplet is given by a 2-form field strength $\cF$
satisfying the modified Bianchi identity $d\cF = -H$. We shall assume
that the non-abelian fields are given by a worldvolume $SU(N)$ 2-form field
strength $F=dA+\frac{1}{\a'}A^2$, $F=F^R t_R$, where the $t_R$ are the generators of the Lie algebra of $SU(N)$. The 2-form $F$
satisfies the standard Bianchi identity,
\be
DF = 0. \label{Bianchi}
\ee
We construct the Lagrangian form from the components of this 2-form
along with the abelian field $\cF^a$ introduced in the previous
section. We shall combine these in such a way as to define a
Lagrangian whose purely bosonic part is a function of
$\hat{F}_{ab}=\cF_{ab}\One + F_{ab}$. For this reason we drop any
terms with fermions in the action but it is possible to calculate
these terms straightforwardly from the equations given below.

In components the Bianchi identity for $F$ (\ref{Bianchi}) reads

\be
\nab_{[A} F_{BC]} + T_{[AB}{}^{D} F_{|D|C]} = 0.
\ee

where $\nab_{A} = \cD_{A} + A_{A}$ is the $SU(N)$ gauge covariant
derivative, with $A_{A}$ being the components of the gauge potential
one form, $A$.

We can take $F_{\a \b} = 0$ without loss of generality by shifting the
bosonic part of the potential. The
solution to the Bianchi identity is then
\begin{align}
F_{\a \b} &= 0, \\
F_{\a b}  &= m^{-1}{}_{b}{}^{c}(\c_{c}\psi)_{\a},  \\
F_{ab}    &= i m^{-1}{}_{a}{}^{c}m^{-1}{}_{b}{}^{d} \e_{cde}(\c^{e})^{\a \b} \nab_{\a} \psi_{\b} + \psi \hspace{3pt} \text{terms}
\end{align}

where we have used the fact that the dimension zero component of the worldvolume torsion is given by
\be
T_{\a \b}{}^{c} = -im_{d}{}^{c}(\c^d)_{\a \b}
\hspace{20pt}
\text{with}
\hspace{20pt}
m_{ab} = (1 + h^2)\eta_{ab} -2 h_a h_b.
\ee

We can use the field $\psi_{\a}=\psi_{\a}^R t_R$ to construct the
closed Lagrangian three form $L_3$ which we require for our action.

The top component of $L_3$, i.e. $L_{abc}$, can be written as
$L_{abc}=\e_{abc}L$. This defines a kappa-invariant action in the same
way as (\ref{abelianS}),

\be
S = \int d^3 x \hspace{3pt} (\det E)  L. \label{action}
\ee 

A shift of the form  $L_{3}\longrightarrow L_{3}+dX_2$ leaves the action $S$ unchanged and this allows us to set $L_{\a\b\c}=0$. We can also use this freedom to set the antisymmetric part of $(\c_b)^{\a\b}L_{\a\b c}$ to zero.

In components $dL_{3}=0$ reads
\be
\cD_{[A} L_{BCD]} + \tfrac{3}{2} T_{[AB}{}^{E}L_{|E|CD]} = 0. \label{Lclosedcomps}
\ee
Using this and the above constraints on $L_3$  we find that 
\be
L_{\a \b c} = i m^{-1}{}_{c}{}^{d}(\c_d)_{\a \b} L_o.
\ee

The idea is then to choose $L_o$ to be a suitable function of the
abelian and non-abelian worldvolume fields and use the closure of
$L_3$ (\ref{Lclosedcomps}) to compute the remaining components of
$L_3$ and hence the action (\ref{action}). 

For now we ignore derivative corrections in the action. We
therefore include no explicit $\a^{\prime}$s in $L_o$.  
Since $L_o$ has dimension $-1$ and the only negative dimension field
on the worldvolume is $\psi_{\a}^R$, with dimension $-\tfrac{1}{2}$,
we can take $L_o$ to be given by the formula 
\be
L_{o} = i \psi_{\a}^R \psi_{\b}^S (J^{o}_{RS} \e^{\a \b} + J^{a}_{RS}
(\c_a)^{\a \b}) + \psi \psi \psi \del \psi ....
\ee
Here $J^{o}_{RS}$,$J^{a}_{RS}$ are functions of the dimension zero
worldvolume fields, $\cF_{ab} , F_{ab}$.

The other components of $L_{3}$ are then given by
(\ref{Lclosedcomps}). We obtain
\begin{align}
L_{\a \b \c} &= 0, \\
L_{\a \b c}  &= i m^{-1}{}_{c}{}^{d} (\c_d)_{\a \b} L_o, \\
L_{\a bc}    &= m^{-1}{}_{b}{}^{d} m^{-1}{}_{c}{}^{e} \e_{dea}
                (\c^a)_{\a}{}^{\b} \cD_{\b} L_o  + \psi \psi \hspace{3pt} \text{terms}, \\
L_{abc}      &= i \e_{abc} (\det m^{-1}) \cD_{\a} \cD^{\a} L_o  + \psi \hspace{3pt} \text{terms}. \label{Leqn}
\end{align}

We see from the final equation that the Lagrangian $L$ is then
determined up to $\psi$ terms by acting with derivatives on $L_o$.
We ignore the $\psi$ terms for now as we are interested in the pure
$\hat{F}$ contribution to the Lagrangian.
The terms which have only $\cF_{ab}=\e_{abc}\cF^c$ and
$F_{ab}=\e_{abc}F^c$ are those where the two derivatives in
(\ref{Leqn}) each act on one of the $\psi$s in $L_o$. To calculate the
Lagrangian the relevant equations are
\begin{align}
L   &= i (\det m^{-1}) \cD_{\a} \cD^{\a} L_o + \psi \hspace{3pt} \text{terms}, \\
F^a &= i (\det m^{-1}) m_{b}{}^{a} (\c^b)^{\a \b} \nab_{\a} \psi_{\b} + \psi \hspace{3pt} \text{terms}.
\end{align}
 
We can absorb the factors of $\det m^{-1}$ into $L_o$ and $\psi$ at
the expense of generating more $\psi$ terms by defining the quantities
\be
\tilde {\psi}_{\a} = (\det m^{-1}) \psi_{\a}
\ee
and
\be
\tL_o = (\det m^{-1}) L_o  = i\tpsi_{\a}^R \tpsi_{\b}^S (\tJ^{o}_{RS} \e^{\a \b} + \tJ^{a}_{RS}(\c_a)^{\a \b})  
+ \tpsi \tpsi \tpsi \del \tpsi ...
\ee

The equations now read
\begin{align}
L   &= i \cD_{\a} \cD^{\a} \tL_o + \tpsi... \label{newLeqn} \\  
F^a m^{-1}{}_{a}{}^{b} &= i (\c^b)^{\a \b} \nab_{\a} \tpsi_{\b} + \tpsi... \label{nabpsi}
\end{align}

Since $\tL_o$ is a gauge scalar we can replace $\cD$ with $\nab$ in
the above equation (\ref{newLeqn}). Employing (\ref{nabpsi}) and noting that the antisymmetric part of $\nab_\a \tpsi_\b$ gives rise to fermion terms we find 
\be
L =  F^{aR} F^{bS} m^{-1}{}_{a}{}^{c} m^{-1}{}_{b}{}^{d} (\eta_{cd}
\tJ^{o}_{RS} - \e_{cde} \tJ^{e}_{RS}) + \tpsi  ... 
\label{newlag}
\ee

The second term involves commutators of $F^a$ since the anti-symmetric
contraction of the Lorentz indices on the $F$s implies antisymmetry of
the gauge indices. We call this term $L_A$. The first term is
symmetric in Lorentz and gauge indices and we call this $L_S$. We shall
combine $L_S$ with pure $\cF$ part of the abelian Lagrangian, $K$,
to give the symmetrised trace term of the non-abelian 
Born-Infeld Lagrangian. The second term, $L_A$, can be used to construct
arbitrary commutator terms.

The full kappa-invariant action for the non-abelian brane is
\be
S^{U(N)} = N S^{U(1)} + S,
\ee
each term being separately kappa symmetric.

Expanding these terms we have
\be
S^{U(N)} = \int d^3 x \hspace{3pt} (\det E) \bigl( N(K - Z) + L_S +
L_A + \psi \text{ terms} \bigr).
\ee

We now explain how the two terms, $L_S$ and $L_A$, can be used to
construct respectively the symmetrised trace part and arbitrary
commutator terms in the non-abelian Born-Infeld Lagrangian.

\subsubsection*{Symmetrised trace Lagrangian}

The first term in (\ref{newlag}) is
\begin{align}
L_S	&= F^{aR} F^{bS} m^{-1}{}_{a}{}^{c} m^{-1}{}_{b}{}^{d} \eta_{cd} \tJ^{o}_{RS} \\
    	&= F^{aR} F^{bS} {1\over(1+h^2)^2}\Bigl( \eta_{ab} + \frac{4h_a h_b}{(1 - h^2)^{2}}\Bigr) \tJ^{o}_{RS} \\
    	&= F^{aR} F^{bS} {1\over(1+h^2)^2}\Bigl( \eta_{ab} + \frac{ \cF_a \cF_b}{1 - \cF^2}\Bigr) \tJ^{o}_{RS} \\
	&= F^{aR} F^{bS} \Bigl( \eta_{ab}(1 - \cF^2) + \cF_a \cF_b
    	\Bigr) A_{RS}. \label{strlag}
\end{align}

In the last line we have absorbed a factor of $(1-h^2)^{-2}$ by defining
$A_{RS}=\tJ^{o}_{RS}(1-h^2)^{-2}$.

To look for a symmetrised trace solution we require
\be
N K + L_S = \Str \sqrt{-\det (\eta_{ab}\One + \hF_{ab})},
\ee
i.e.
\be
N\sqrt{1 - \cF^2} + \Str \Biggl( F^a F^b \Bigl( \eta_{ab} (1 - \cF^2) + \cF_a \cF_b \Bigr)\bbA \Biggr)
= \Str \sqrt{\One - (\cF_a \One + F_a) (\cF^a \One +  F^a)}.
\label{strlageqn}
\ee
The first term on the left hand side is the purely abelian
contribution, the second being the new non-abelian parts of the
Lagrangian. $\bbA$ is an $N\xz N$ matrix which will be a power series in the $SU(N)$ generators $t_R$. We have

\be
\bbA=A_0\One + A^S t_S + A^{ST} t_S t_T + \ldots
\ee

where each coefficient function is symmetric. The tensor $A_{RS}$ is then given by

\be
A_{RS}=\d_{RS}A_0 + d_{RST} A^T + d_{RSTU} A^{TU} + \ldots
\ee

Equation \eq{strlageqn} can now be solved for $\bbA$. We introduce the
variables 
\be
X = \cF_a \cF^a \One, \hspace{20pt} Y = \cF_a F^a, \hspace{20pt} Z = F_a F^a.
\ee

The equation now reads 
\be
\Str \biggl( \sqrt{\One - X} + \Bigl( Z(\One - X) + Y^2 \Bigr) \bbA \biggr) = \Str \sqrt{\One - (X + 2Y + Z)}.
\ee

If we expand out the square root on the right hand side we notice that
the trace will kill terms which are linear in $Y$ and have no
$Z$. These terms can be written explicitly as
\be
-Y(\One - X)^{-\tfrac{1}{2}}.
\ee

If we explicitly remove these terms from the equation we find that $\bbA$ is given by
\be
\sqrt{\One - X} + \Bigl( Z(\One - X) + Y^2 \Bigr) \bbA =  \sqrt{\One - (X + 2Y + Z)} + Y(\One - X)^{-\tfrac{1}{2}}.
\ee

We do not have to worry about the ordering of the non-abelian
quantities $Y$ and $Z$ because under the symmetric trace operation
everything effectively commutes. This equation defines $\bbA$
as a Taylor expansion in $X,Y,Z$ about $X=0, Y=0 ,Z=0$, 
\be
\bbA = \sum_{l,m,n = 0}^{\infty} a_{l,m,n}X^l Y^m Z^n.
\ee
Note that this
would not be the case if the relative value of the coefficients in 
equation (\ref{strlageqn}) were different. The existence of a
non-singular solution to this equation shows that this construction
can be used to obtain a supersymmetric, kappa-invariant action whose
pure $\hF$ terms give the symmetrised trace non-abelian Born-Infeld
Lagrangian. The first few terms in the Taylor expansion of $\bbA$ are

\be
\begin{aligned} 
\bbA = &-\tfrac{1}{2}\One - \tfrac{3}{4}X -\tfrac{1}{2}Y -\tfrac{1}{8}Z  -\tfrac{15}{16}X^2 -\tfrac{5}{4}XY \\
    &-\tfrac{5}{8}Y^2 -\tfrac{5}{16}XZ -\tfrac{3}{8}YZ -\tfrac{1}{16}Z^2 + .... 
\end{aligned}
\ee

This solution corresponds to the following expression for $A_{RS}$ in equation (\ref{strlag}):

\be
\begin{aligned}
A_{RS} = -{1\over2} &\left\{
\d_{RS}\left(1+{3\over2}\cF^2 +{15\over8}\cF^4+\ldots\right)\right.\\
&+d_{RST}\left(\cF_{a} F^{aT}+{5\over2}\cF^2 \cF_{a} F^{aT}+\ldots\right)\\
&+d_{RSTU}\left({1\over4}F_{a}^{T}F^{aU}+{5\over4}\cF_{a} F^{aT} \cF_{b} F^{bU}+{5\over8}\cF^2 F_{a}^{T}F^{aU}+\ldots\right)\\
&+d_{RSTUV}\left({3\over4}\cF_{a}F^{aT} F_{b}^{U}F^{bV}+\ldots\right)\\
&\left.+d_{RSTUVW}\left({1\over8}F_{a}^{T}F^{aU}F_{b}^{V}F^{bW}+\ldots\right)
+\ldots\right\}
\end{aligned}
\ee

where $d_{RST} = \Str (t_{R}t_{S}t_{T})$ etc. are the $SU(N)$ d-symbols.

\subsubsection*{Commutator terms}

To obtain commutator terms we use the second term in equation
(\ref{newlag}),
\be
L_A = - F^{aR} F^{bS} m^{-1}{}_{a}{}^{c} m^{-1}{}_{b}{}^{d} \e_{cde} \tJ^{e}_{RS}.
\ee
If we define $A^{a}_{RS}$ by $\tJ^{a}_{RS} = (\det m)
m^{-1}{}_{b}{}^{a} A^{b}_{RS}$ then we find 
\be
L_A = - F^{aR} F^{bS} \e_{abc} A^{c}_{RS}. \label{commlag}
\ee 

We can now choose $A^{c}_{RS}$ to be any function of the $U(N)$ field
strength $\hF^a = \cF^a + F^a \One$ so as to incorporate any
desired commutator terms in the Lagrangian. For example at order
$\hF^3$ in the Lagrangian we can have
\be
A^{c}_{RS} = - F^{cT} f_{RST},
\ee
which gives a term in the Lagrangian of the form,
\be
\Tr F^a F^b F^c \e_{abc} = \Tr \hF^a \hF^b \hF^c \e_{abc}.
\ee

At order $\hF^4$ in the Lagrangian we can have
\be
A^{c}_{RS} = - \e^{cde}F_{d}^{U}F_{e}^{V} f_{RST}f^{T}{}_{UV},
\ee
which gives a term in the Lagrangian of the following type,
\be
\Tr [F^{a},F^{b}][F_{a},F_{b}] = \Tr [\hF^{a},\hF^{b}][\hF_{a},\hF_{b}].
\ee

\subsubsection*{Full action}

In summary the full action we have constructed is as follows:

\be
S^{U(N)} = \int d^3 x (\det E)(NL^{U(1)} + L_S + L_A + \psi \text{ terms}).
\ee

We have shown that $L_S$ can be chosen such that the purely bosonic parts of the first and second terms combine to give the symmetrised
trace Lagrangian and that $L_A$ can give any
choice of commutator terms for $\hF$ including those with odd powers
of $\hF$. We therefore have

\be
S^{U(N)} = \int d^3 x (\det E) \biggl( \Str \sqrt{-\det(\eta_{ab} \One + \hF_{ab})}
- N Z  + \text{ commutator terms } + \psi \text{ terms} \biggr),
\label{fullaction}
\ee

where the superfield integrand is evaluated at $\th=0$ as usual.
The kappa-symmetry of this action is guaranteed by the fact that it is
constructed from the sum of two closed forms (the abelian one and the
new one).


\section{Conclusions}

In this paper we have shown how to construct a manifestly kappa-symmetric non-abelian action for the space-filling brane in three
dimensions. The invariance under a single kappa-symmetry
is equivalent to the local worldvolume supersymmetry of
the system in the superembedding picture. 

The basic idea is to extend the abelian action by adding a new invariant involving an $SU(N)$ worldvolume gauge supermultiplet in such a way that the resulting action is a single trace over a
function of the $U(N)$ field strength. The $SU(N)$ field strength multiplet is described by a spinorial superfield $\psi$
of  dimension $-1/2$ which allows one to construct a closed
Lagrangian 3-form from a scalar superfield, $L_o$, of dimension $-1$. Note that, in three dimensions, the Wess-Zumino term is the same as in the abelian case. This means that the $SU(N)$ part of the non-abelian Born-Infeld action comes entirely from $L_o$ and is not determined by the WZ term. In higher dimensions the latter will have non-abelian contributions and this will necessitate a slightly modified approach to the problem. 

We note that kappa symmetry (and target space supersymmetry) does not determine the non-abelian action uniquely, at least in the model under consideration. The symmetrised trace contribution seems to be fixed but it is possible to add many different commutator terms as we have seen. For each of these actions our method guarantees supersymmetry and the fermion contributions could be worked out straightforwardly. However, we have not worked these out in detail. It might be that the structure of these terms could imply further restrictions on the form of the action.

If we include derivative terms in
the functions $A_{RS}$ of equation (\ref{strlag}) and $A^{a}_{RS}$ of
equation (\ref{commlag}) we can produce derivative corrections in
the Lagrangian. Such terms would be invariant independently of the action constructed above in equation (\ref{fullaction})and would in some sense be analogous to those found in \cite{hl} using a similar procedure.

It should be easy to generalise the discussion in this paper to a 2-brane embedded in a curved $N=2, D=3$ supergravity background. This would modify the background curvatures but would not substantially alter the procedure for obtaining the non-abelian Born-Infeld action, althought there would clearly be couplings to the supergravity fields. We could also look for terms involving higher derivatives of the background curvature \cite{Bachas:1999um}. However, the dimension of spacetime is too low  for there to be background curvature corrections to the Wess-Zumino term of the type found in \cite{Green:1996dd,Cheung:1997az}.

It should also be possible to dimensionally reduce the action \eq{fullaction} in different ways. For example, reducing the worldvolume to a worldline would give an action describing a D-particle moving in three spacetime dimensions while double dimensional reduction followed by a reduction of the worldvolume to a worldline would allow a comparison with the results of \cite{sorokin}.

\vskip .5cm
\section*{Acknowledgements}
This article represents work carried out under EU contract HPNR-CT-2000-0122 and which was also supported in part by PPARC through SPG grant 68 and by VR grant 5102-20005711.


\end{document}